\documentclass[superscriptaddress,aps,prX,twocolumn,showpacs,nofootinbib,longbibliography,notitlepage,floatfix]{revtex4-2}
\usepackage[latin9]{inputenc}
\usepackage{mathtools}
\usepackage{amsbsy}
\usepackage{amstext}
\usepackage{xcolor}
\usepackage{upgreek}

\usepackage[unicode=true,pdfusetitle,
 bookmarks=false,
 breaklinks=false,pdfborder={0 0 1},backref=false,colorlinks=false]
 {hyperref}

\usepackage{etoolbox}

\usepackage[english]{babel}
\usepackage{chngcntr}
\hypersetup{
 bookmarksnumbered=false,bookmarksopen=false,colorlinks,citecolor=blue,urlcolor=blue,filecolor=blue,linkcolor=blue}
\makeatletter

\@ifundefined{textcolor}{}
{
 \definecolor{BLACK}{gray}{0}
 \definecolor{WHITE}{gray}{1}
 \definecolor{RED}{rgb}{1,0,0}
 \definecolor{BLUE}{rgb}{0,0,1}
 \definecolor{CYAN}{cmyk}{1,0,0,0}
 \definecolor{MAGENTA}{cmyk}{0,1,0,0}
 \definecolor{YELLOW}{cmyk}{0,0,1,0}
 \definecolor{ORANGE}{rgb}{1,0.5,0}
}

\usepackage{amsmath}
\usepackage{graphicx}
\usepackage{amssymb}
\usepackage{txfonts,color}
\makeatother

\begin{document}

\title{Amplified Nanoscale Detection of Labelled Molecules via Surface Electrons on Diamond}
\author{Ainitze Biteri-Uribarren}
\thanks{These authors have equally contributed to this work.}
\affiliation{Department of Physical Chemistry, University of the Basque Country UPV/EHU, Apartado 644, 48080 Bilbao, Spain}
\affiliation{EHU Quantum Center, University of the Basque Country UPV/EHU, Leioa, Spain}

\author{Pol Alsina-Bol\'{i}var}
\thanks{These authors have equally contributed to this work.}
\affiliation{Department of Physical Chemistry, University of the Basque Country UPV/EHU, Apartado 644, 48080 Bilbao, Spain}
\affiliation{EHU Quantum Center, University of the Basque Country UPV/EHU, Leioa, Spain}

\author{Carlos Munuera-Javaloy}
\affiliation{Department of Physical Chemistry, University of the Basque Country UPV/EHU, Apartado 644, 48080 Bilbao, Spain}
\affiliation{EHU Quantum Center, University of the Basque Country UPV/EHU, Leioa, Spain}

\author{Ricardo Puebla}
\affiliation{Department of Physics, University Carlos III of Madrid, Avda. de la Universidad 30, 28911  Legan{\'e}s, Madrid, Spain}

\author{Jorge Casanova}
\affiliation{Department of Physical Chemistry, University of the Basque Country UPV/EHU, Apartado 644, 48080 Bilbao, Spain}
\affiliation{EHU Quantum Center, University of the Basque Country UPV/EHU, Leioa, Spain}
\affiliation{IKERBASQUE,  Basque  Foundation  for  Science, Plaza Euskadi 5, 48009 Bilbao,  Spain}
\affiliation{Corresponding author: jcasanovamar@gmail.com}

\begin{abstract}
{\bf ABSTRACT:} The detection of individual molecules and their dynamics is a long-standing challenge in the field of nanotechnology. In this work, we present a method that utilizes a nitrogen vacancy (NV) center and a dangling bond on the diamond surface to measure the coupling between two electronic targets tagged on a macromolecule. To achieve this, we design a multi-tone dynamical decoupling sequence that leverages the strong interaction between the nitrogen vacancy center and the dangling bond. In addition, this sequence minimizes the impact of decoherence finally resulting in an increased signal-to-noise ratio. This proposal has the potential to open up avenues for fundamental research and technological innovation in distinct areas such as biophysics and biochemistry.
\end{abstract}
\maketitle

\section{Introduction}
Finding the structural disposition and dynamics of macromolecules is central in life sciences. For instance, in biology or biochemistry, it is of great interest to track the conformational changes of proteins to understand biological processes and treat protein miss-folding related diseases \cite{protein}. In this scenario, nuclear magnetic resonance (NMR) techniques provide a valuable analytical tool for researchers, allowing them to investigate the conformational characteristics and intermolecular associations of biomolecules~\cite{marion2013introduction}.  However, standard NMR spectroscopy~\cite{abragam1961principles, levitt2013spin}, magnetic resonance imaging~\cite{weishaupt2006how} or electron-spin resonance~\cite{prisner2001pulsed, schweiger2001principles} are limited by their inherent low sensitivity, making them applicable primarily to bulky samples. In other words, due to the low thermal nuclear polarization achievable at room temperature, a sample with a volume greater than a hundred microliters \cite{allert2022advances} is typically required to generate detectable responses. As a result, standard NMR-based techniques are unsuitable for studying nanoscopic- and microscopic-sized samples.

The nitrogen vacancy (NV) center \cite{doherty2013nitrogen}, a promising single-spin quantum sensor, is gaining popularity in this context. It is a spin-one system that offers long coherence times \cite{allert2022advances,abobeih2018one} even at room temperature \cite{ohashi2013negatively,lovchinsky2016nuclear}, it can be initialized and readout using a green laser~\cite{steiner2010universal,waldherr2011dark,cai2013diamond} while its hyperfine levels can be easily manipulated with microwave (MW) radiation \cite{tamarat2008spin,allert2022advances,Neumann2008}.
Due to its atomic size, the NV center can be placed in close proximity to the target sample, thereby increasing their coupling which results in a large sensitivity and spatial resolution~\cite{abobeih2019atomic,shi2015single, schlipf2017molecular}. Owing to these unique properties, the NV center is used to  perform NMR experiments with micrometer~\cite{glenn2018high, arunkumar2021micron} and nanometer~\cite{bucher2020hyperpolarization,mamin2013nanoscale,staudacher2013nuclear} resolution, as well as electron-spin resonance experiments~\cite{meinel2021heterodyne,staudenmaier2021phase}.

A potential use of NV-based sensing is the measurement of the distance between specific sites in a single protein, which can help to track its folding process. This could be achieved, e.g., by attaching a pair of electron spin labels (in the following called \textit{labels}) to key sites of the target protein and measuring the {\it inter-label coupling constant} before and after the folding process~\cite{munuera2022detection}. To achieve this, the labeled protein should be placed on the surface of a diamond close to a shallow NV defect (a few nanometers). It is worth noting that previous studies have demonstrated NV detection of single nitroxide labels~\cite{shi2015single}. Additionally, it has even demonstrated the detection of inter-label coupling in synthetic peptides~\cite{schlipf2017molecular}, highlighting the potential of NV-based sensing for structural and dynamical studies. While current methods have shown promising results in using NVs to detect single labels and inter-label coupling, there is still ample opportunity for improvement in terms of sensitivity and signal-to-noise ratio (SNR). Hence, incorporating hybrid sensors and advanced quantum control techniques has the potential to significantly enhance NV-based sensing capabilities.

In this work, we introduce a detection protocol that leverages a hybrid sensor consisting of a shallow NV and a dangling bond (DB); the latter is an unpaired immobilized electron that tends to appear on the diamond surface. Even though DBs are often regarded as a noise source in NV sensing, our protocol takes advantage of their presence as mediators that enhance the detection of the coupling between two labels attached to a molecule. To reach this goal we design a multi-tone dynamical decoupling sequence that encodes the inter-label coupling constant in oscillations of the NV fluorescence through a DB mediator, while minimizes the effects of decoherence on both NV and DB. Through detailed numerical simulations, we demonstrate the enhanced signal-to-noise ratio achieved by our method compared to the standard scenario without DB. Overall, our work has the potential to open up new avenues for fundamental research and technological innovation by using nanoscale hybrid sensors.

\section{Results and discussion}
\subsection{The system}

A scheme of the considered scenario is in Fig.~\ref{seq}(a) This encompasses a shallow NV, a surface electron spin (i.e. a DB) and two labels ($\rm{L_1}$ and $\rm{L_2}$). The system Hamiltonian is:

\begin{align}
H/ \hbar &= D \left(S^{\rm{z}}\right)^2 +B^{\rm{z}}|\gamma_{\rm{e}}| S^{\rm{z}} + B^{\rm{z}} |\gamma_{\rm{e}}| J_{\rm{z}} + H_{\rm{ L_1}}+H_{\rm {L_2}}\nonumber\\[8pt] 
&+H^{\rm{dd}}_{\rm {NV-DB}} + \sum_{\rm{j=1,2}} H^{\rm{dd}}_{\rm{DB-L_j}} + H^{\rm{dd}}_{\rm {L_1-L_2}} + \sum_{\rm{j=1,2}}H^{\rm{dd}}_{\rm{NV-L_j}} \nonumber\\[8pt]
&+ H_{\rm{c}}. \label{eq:SfullHamiltonian}
\end{align}

Here, $D\approx (2\uppi)\times 2.87$ GHz is the zero-field splitting of the NV, $|\gamma_{\rm{e}}|=(2\uppi)\times 28\cdot10^3\:\rm{MHz\:T^{-1}}$ is the electronic gyromagnetic ratio and $B^{\rm{z}}$ is the external field, which is aligned with the NV axis (here the $z$-axis), leading to the Zeeman splitting $B^{\rm{z}}|\gamma_{\rm{e}}| S^{\rm{z}}$ with $S^{\rm{z}}$ being the NV spin operator. In addition, $J_{\rm{z}}$ is the spin operator of the DB and $B^{\rm{z}} |\gamma_{\rm{e}}| J_{\rm{z}}$, $H_{\rm L_1}$, $H_{\rm L_2}$ are the free-energy terms of the dangling bond and labels, respectively. Effectively, $H_{\rm{L_i}}$ can be reduced to $H_{\rm{L_i}}\approx \omega_{\rm{i}} P_{\rm{i}}^{\rm{z}}$, where $P_{\rm{i}}$ is the spin operator of the i-th label and $\omega_{\rm{i}}$ is its corresponding resonance energy~\cite{shi2015single,schlipf2017molecular,munuera2022detection}. The terms on the second line hold for the dipole-dipole interaction between every pair of system constituents, see Fig.~\ref{seq}(a), which for two arbitrary electronic elements i and j, with spin operators $S_{\rm{i}}$ and $S_{\rm {j}}$, reads as $H^{\rm{dd}}_{\rm {i-j}}  = \frac{\mu_0 \gamma_{\rm{e}}^2 \hbar}{4 {\rm\uppi} d_{\rm{ij}}^3}\left[{\bf S_{\rm{i}}}\cdot{\bf S_{\rm{j}}}-\frac{3\left({\bf S_{\rm{i}}}\cdot{\bf r_{\rm{ij}}}\right)\left({\bf S_{\rm{j}}}\cdot{\bf r_{\rm{ij}}}\right)}{d_{\rm{ij}}^2}\right]$. The last term in Eq.~(\ref{eq:SfullHamiltonian}) is the control Hamiltonian $H_{\rm c}$, and encompasses the MW driving fields.

We assume that the system constituents --i,e, NV, DB and labels-- have different resonance energies. We will now provide a justification for this assumption. On the one hand, the resonance energy of the NV naturally differs from that of the other electronic character members due to zero field splitting $D$. On the other hand, the labels tend to exist in specific molecular environments which inherently shift their resonance energy up to hundreds of MHz. In particular, for labels encoded in nitroxide-based radicals, their different orientations with respect to (w.r.t.) the magnetic field $B^{\rm z}$ led to energy differences which have  been utilised to selectively address individual labels~\cite{schlipf2017molecular}. Alternatively, one could employ a magnetic tip to ensure different resonance energies, this generates a magnetic field gradient that causes varying Zeeman energy splittings. In this respect, gradients of $\approx 60\:{\rm G\: nm^{-1}}$ have been already reported using FeCo tips~\cite{mamin2012high}. Then, owing to the differences among the resonance energies, Eq.~(\ref{eq:SfullHamiltonian}) greatly simplifies by dropping all the transversal terms (i.e. those that do not contain the z component of spin operators) leading to only ZZ-type interactions as Eq. (\ref{eq:ZZ}) depicts (from now on, we will refer to the interaction between the \emph{z} components of two spins as \emph{ZZ interactions}). For that, one needs to move to a rotating frame w.r.t. the free energy terms of NV, DB and labels, and invoke the rotating wave approximation (RWA) to eliminate fast rotating terms, which results in

\begin{align}
H/\hbar =&  A^{\rm z}_{\rm NV-DB} S_{\rm z}J_{\rm z} + \sum_{\rm j=1,2} A^{\rm z}_{\rm DB-L_j} J_{\rm z} P^{\rm z}_{\rm j} + \sum_{\rm j=1,2} A^{\rm z}_{\rm NV-L_j} S_{\rm z} P^{\rm z}_{\rm j}  \nonumber\\[6pt] 
+& A^{\rm z}_{\rm L_1-L_2} P^{\rm z}_1 P^{\rm z}_2 + \tilde{H}_{\rm c},\label{eq:ZZ}
\end{align}
where $A^{\rm z}_{\rm i-j} =\frac{\mu_0 \gamma_{\rm e}^2 \hbar}{4 {\rm \uppi} d_{\rm ij}^3} \left[1 -3 \cos^2{(\theta_{\rm i-j})}\right]$ is the amplitude of the remaining ZZ dipole contribution between the i-th and j-th elements after the RWA has been applied. Here, $\theta_{\rm i-j}$ is the angle between the joining vector of the i-th and j-th elements and the direction of the magnetic field (${\bf B}=B^{\rm z}{\bf\hat{\rm z}}$) (for more details, see Supplementary Note 1). From this point on, to facilitate the notation, we use $g$ to refer to the inter-label dipolar coupling constant (i.e. $g \equiv A^{\rm z}_{\rm L_1-L_2}$).

\subsection{The protocol}

\begin{figure*}[]
\includegraphics[width= 1 \linewidth]{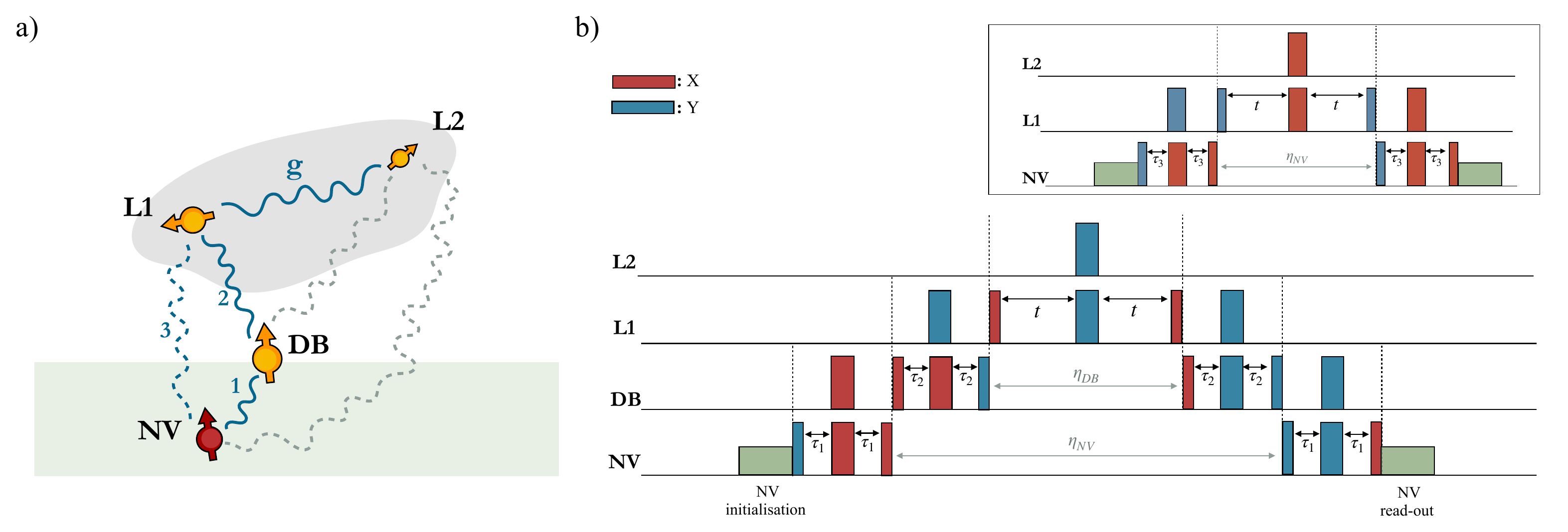}
\caption{{\bf Scheme of the system and the designed radiation pattern.} (a)Schematic representation of the system. The dipolar coupling $g$ between the labels is the target parameter of the protocol. Interactions 1 and 2 are used in the protocol involving the dangling bond (DB), whereas connection 3, between the nitrogen vacancy (NV) and label 1 ($\rm L_1$)   is employed in the scheme with no DB. The nonutile interactions in each case are canceled due to the dynamical decoupling nature of both sequences. (b) Scheme of the multi-tone dynamical decoupling sequence, where each channel is associated to an element of the system (NV, DB, and labels). The blue and red boxes represent the different pulses to be delivered, the wider ones indicate $\pi$ pulses while the others $\uppi/2$ pulses, and the color encodes the axis of the pulse. Besides, the duration of each spin-echo is ($2\tau_1$, $2\tau_2$ and $2t$). Inset: the alternative protocol to be executed when there is no DB. The NV interacts directly with ${\rm L_1}$ during the time interval $2\tau_3$.}\label{seq}
\end{figure*}

Fig.~\ref{seq} (b) illustrates our sequence, which involves ${\uppi}$ and ${\rm \uppi/2}$ pulses along distinct axes applied to the NV, DB and labels. These pulses are arranged in a castle-shaped scheme of spin echoes that encodes the coupling between the labels ($g$) into the final quantum state of the NV. Our protocol, inspired by correlation spectroscopy methods~\cite{laraoui2013high}, minimises the effect of decoherence by reducing the time in which the elements are exposed to dephasing. In particular, the dephasing that suffer the system constituents only comes into play inevitably during the short double electron-electron resonance (DEER) blocks of length $\tau_1$, $\tau_2$, and $t$ (see Fig.~\ref{seq} (b)) that communicate  NV with DB, DB with the first label, and the first label with the second label, respectively; and even during these periods, the DEER structure cancels the undesired interactions with the other elements. Other effects of dephasing, such as the ones induced by random phases on the NV and DB  (i.e., $\eta_{\rm NV}$ and $\eta_{\rm DB}$) accumulated during the long periods indicated in Fig.~\ref{seq} (b), can be suppressed. More specifically, our protocol confines the additional contribution of these random phases in terms (second line of Eq.~(\ref{noise1})) distinct to that we intend to evaluate (first line of Eq.~(\ref{noise1})) which includes the target parameter $g$.

For the sake of simplicity in the analytical calculation, we assume that $\tilde{H}_{\rm c}$ (see Eq.~(\ref{eq:ZZ})) delivers individual controls in the form of $\uppi$ and $\uppi/2$ pulses to each system constituent (however, our numerical simulations include finite-width pulses and crosstalk effects among NV, DB, and labels). Besides, we consider the DB and labels in a thermal state while the NV is initialized to the ground state. In this case, applying the sequence pictured in Fig.~\ref{seq} (b) leads to the following outcome for the NV population:

\begin{align}\label{noise1}
    P_0&= \frac{1}{2}\left[1-\sin^2(\phi_1)  \sin^2 (\phi_2) \cos \left(g t \right)\right.\nonumber\\[6pt]
    &\left.+\cos^2(\phi_1)\sin(\eta_{NV})-\sin^2(\phi_1)\cos^2(\phi_2)\sin(\eta_{DB})\right],
\end{align}
where $P_0\equiv\rm{Tr\left(\rho\:|0\rangle\langle0|_{\rm{NV}}\right)}$ is the probability of finding the NV in the ground state and $\phi_1=\left(A_{\rm{NV-DB}}^{\rm z}\right)\tau_1$ and $\phi_2=\left(A^{\rm z}_{\rm{DB-L_1}}\right)\tau_2$ are the accumulated phases during the intervals $2\tau_1$ and $2\tau_2$, respectively. Thus, the noise due to $\eta_{\rm{NV}}$ and $\eta_{\rm{DB}}$ can be set to zero by suitably fixing $\tau_1$ and $\tau_2$ such that $\phi_1=\phi_2=\uppi/2$ leading to 

\begin{align}\label{unitary}
    P_0 = \frac{1-\cos \left(g t \right)}{2} .
\end{align}

In an experimental setup, where a priori the coupling amplitudes are unknown, to meet this condition for optimal performance (i.e. $\phi_1=\phi_2=\uppi/2$), one could simply conduct different experiments with varying $\tau_1$ and $\tau_2$, while tracking the amplitude of the harvested signal in order to find a better estimation of $g$. Furthermore, in the event of an incorrect attainment of the phases $\phi_1=\phi_2=\uppi/2$, the stochastic nature of the noise tends to average out their effect when the NV is interrogated several times. This is, $\frac{1}{N}\sum_{\rm j=1}^N \sin(\alpha_{\rm j}) \rightarrow 0$, with $N$ being the number of experimental acquisitions, and $\alpha_j$ the accumulated phase over the NV (or DB) owing to random dephasing at the j-th interrogation, leading, also this way, to  Eq.~(\ref{unitary}), where the term containing $g$ is untouched.

In summary: Our approach directly encodes the coupling between the labels in the oscillation frequency of the NV population and diminishes considerably the effect of the dephasings. In addition, it eliminates the undesired interactions among system constituents during the communication periods $\tau_{\rm i}$ due to the DEER structure. This allows capturing a finite number of oscillations under realistic conditions, as shown in the next section. 

If we consider a scenario without the DB our scheme simplifies to that in Fig.~\ref{seq}.(b) inset leading to

\begin{align}\label{noiseNV}
  	 P_0  =\frac{1}{2}\left[1 -\sin^2(\phi_3) \cos \left(g t \right) +\cos^2(\phi_3)\sin(\eta_{\rm NV})\right],
\end{align}
where $\phi_3=(A_{\rm NV-SL1}^z)\tau_3$.
This alternative scheme, which does not involve a DB, is two spin-echo periods shorter. However, due to the large distance between the NV and the labels, their coupling is significantly reduced (it is important to note the $\frac{1}{d_{\rm j}^3}$ scaling of the NV-labels interaction in this regard). Hence, in order to have  $\phi_3=(A_{\rm NV-SL1}^z)\tau_3=\uppi/2$, $\tau_3$ needs to be much longer than in the protocol with DB.

\subsection{Numerical results}
We conducted simulations on two scenarios: $(i)$ Utilizing a dangling bond as a signal amplifier with the sequence depicted in Fig.~\ref{seq}(b), and $(ii)$ In the absence of a dangling bond, employing the pulses shown in the inset of Fig.~\ref{seq}(b). In this section, we compare the results obtained from these simulations
.
The influence of decoherence channels (dephasing and thermalization) is accounted for through the implementation of a master equation in Lindblad form.
In particular we use $T_{\rm 2,NV}=5 \ \mu$s, $T_{\rm 2,DB}=1 \ \mu$s, $T_{\rm 1,NV}=20  \mu$s and $T_{\rm 1,DB}=29.4  \mu$s~\cite{sushkov2014magnetic}, while for the labels we take $T_{\rm 1, L_1}=T_{\rm 1, L_2}=4  \mu$s and $T_{\rm 2, L_1}=T_{\rm 2, L_2}=1\ \mu$s~\cite{shi2015single} . Refer to "Implementation of decoherence" subsection in Methods for further details. 

The performance of both protocols highly depends on the relative position of the elements. The dipolar coupling is inversely proportional to the cube of the distance; besides, the interaction between every two elements is strongly dependent on the angle between their joining vector and the direction of the applied magnetic field. Indeed, at the so-called magic angle ($\theta_{\rm magic} = 54.7^{\circ}$), the spin-spin interaction is nullified, resulting in a strong reduction of dipole coupling as the angle approaches $\theta_{\rm magic}$ (see expression for the ZZ interaction in Eq.~(\ref{eq:ZZ})). Hence, this must be taken into account to perform the comparison. In this case, we consider that an appropriate scenario is that in which the angles of all the interactions are similar.
In particular, we choose a configuration of NV, DB, and labels such that: $d=5.6$ nm (NV-DB), $d_1=11.3$ nm (NV-$\rm{L_1}$), $d_2=14.8$ nm (NV-$\rm{L_2}$), $d_{\rm DB-L_1}=5.7$ nm (DB-$\rm{L_1}$), $d_{\rm DB-L_2}=9.3$ nm  (DB-$\rm{L_2}$) and $d_{12}=3.8$ nm  ($\rm{L_1}$-$\rm{L_2}$). As for the angles: $\theta_{\rm NV-DB}=13.8^{\circ}$, $\theta_{\rm NV-L_1}=13.4^{\circ}$, $\theta_{\rm DB-L_1}=13.5^{\circ}$ (see Supplementary Note 1 for further details). This gives rise to the following coupling constants: $A^{\rm z}_{\rm NV-DB}=(2\uppi)\times 0.550 \:\rm{MHz}$, $A^{\rm z}_{\rm NV-L_1}=(2\uppi)\times 66 \:\rm{kHz}$, $A^{\rm z}_{\rm NV-L_2}=(2\uppi)\times 32 \:\rm{kHz}$, $A^{\rm z}_{\rm DB-L_1}=(2\uppi)\times 0.511\: \rm{MHz}$, $A^{\rm z}_{\rm DB-L_2}=(2\uppi)\times 0.130 \:\rm{MHz}$, $g\equiv A^{\rm z}_{\rm L_1-L_2}=(2\uppi)\times 1.734 \:\rm{MHz}$. Using this setting, the protocol involving the hybrid NV-DB sensor has a duration of $10 \ \mu$s, whereas the protocol without a DB lasts $21 \ \mu$s. 

\begin{figure}[t]
\includegraphics[width=\linewidth]{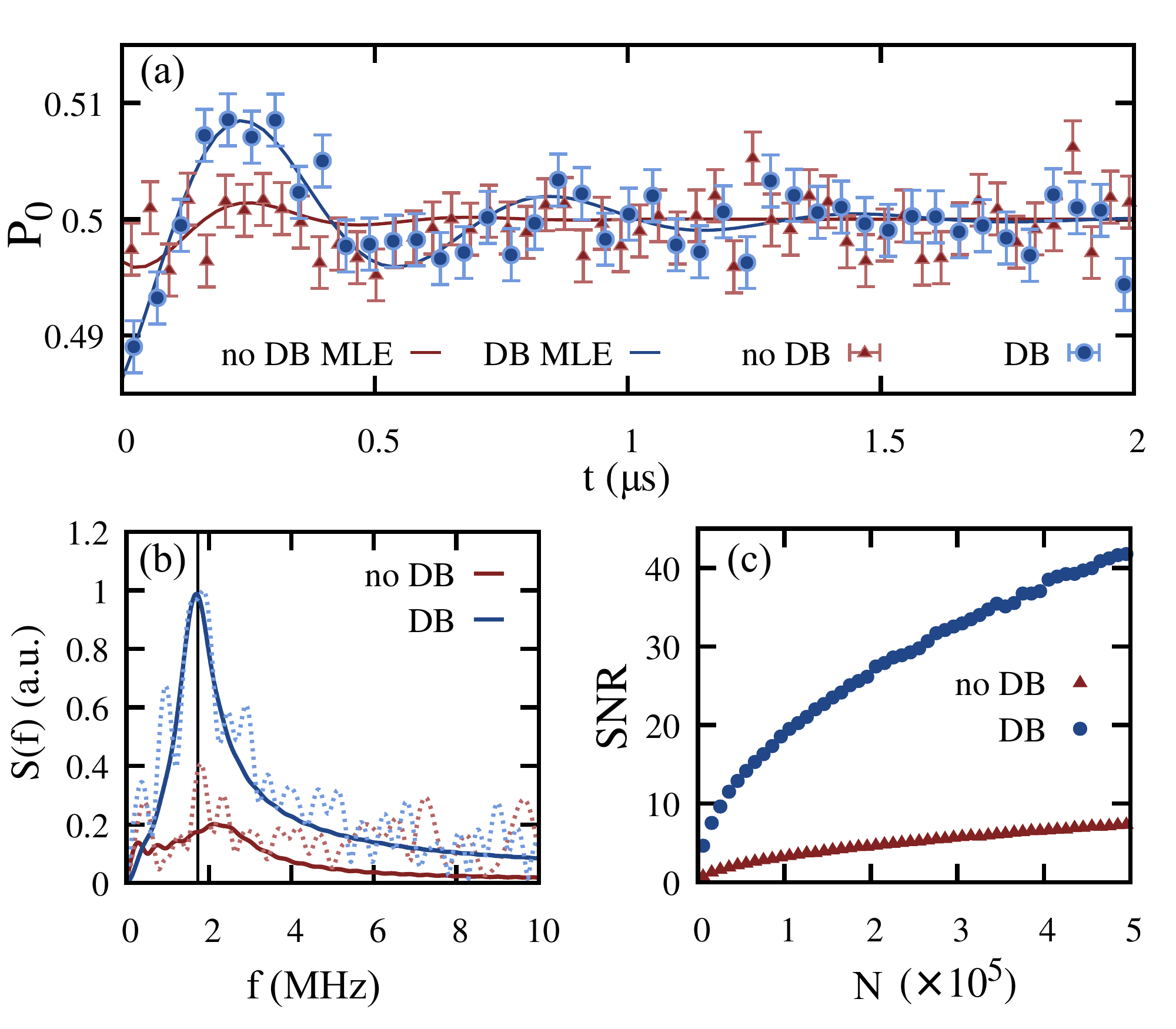}
\caption{{\bf Simulated signals and corresponding spectra. }(a) Signal obtained with the hybrid NV-DB sensor (in blue) and with the single NV (in red). Each obtained value (blue circle or red triangle) has been computed with 50000 shots where bars show the corresponding statistical noise, while solid lines correspond to the maximum-likelihood estimate (see the ``Maximum likelihood estimation'' subsection in Methods). (b) Rescaled power spectrum $S(f)$ of the Fourier transform applied to the data in (a). The results corresponding to the hybrid sensor are represented in blue. Specifically, the blue line is the Fourier transform of the blue fit, while the dotted blue curve corresponds to the blue circles including statistical noise from (a). Solid and dotted red lines depict the equivalent quantities without DB. Note this last case is fully covered by noise. A vertical line is drawn at the true value $g\equiv A^{\rm z}_{\rm L_i-L_j} =\frac{\mu_0 \gamma_{\rm e}^2 \hbar}{4 {\rm \uppi} d_{\rm 12}^3} \left[1 -3 \cos^2{(\theta_{\rm L_1-L_2})}\right]=1.734$ MHz, around which the obtained power spectrum is centered. (c) Depicts the SNR of both methods as a function of the number of experiments~N.}
\label{db_vs_nodb}
\end{figure}

We take into account the  occurrence of cross-talk effects resulting from imperfect individual addressing when the system is subjected to MW radiation (see the ``Cross-talk'' subsection in Methods). In this context, it is important to note that the resonance frequency of the NV is protected by the zero field splitting $D$, while the presence of a magnetic field gradient of $\approx30\:{\rm G\: nm^{-1}}$ (note this gradient is half the value reported in experimental studies \cite{mamin2012high}) separates the transition frequencies of the DB and labels thus improving single addressability. The external static magnetic field can be chosen in a wide range (values on the order of 150 G are common in NV-based setups \cite{shi2015single} and \cite{laraoui2013high}) as long as one reaches Eq.~(\ref{eq:ZZ}) where fast rotating terms cancel, while the introduced gradient enables addressing of the distinct system constituents. In these conditions the transition frequencies corresponding to the free terms of Eq.~(\ref{eq:SfullHamiltonian}) are $w_{\rm{NV}}\equiv D+|\gamma_{\rm e}|B^{\rm z}({\bf r}_{\rm{NV}})=(2\uppi)\times 3.29\:\rm{GHz}$,  $w_{\rm{DB}}\equiv |\gamma_{\rm e}|B^{\rm z}({\bf r}_{\rm{DB}})=(2\uppi)\times 0.826\:\rm{GHz}$,  $w_{\rm{SL1}}=(2\uppi)\times 1.24\:\rm{GHz}$ and  $w_{\rm{SL2}}=(2\uppi)\times 1.55\:\rm{GHz}$. With this choice and a driving amplitude of $\Omega=(2\uppi)\times 10\text{ MHz}$ leading to $\uppi$ ($\uppi/2$) pulses of 50 ns (25 ns), crosstalk effects are visible but do not disturb the estimation of $g$. As explained in the ``Implementation of Decoherence'' subsection in Methods, our numerical simulations are conducted taking the Eq.~\ref{eq:Linbland} as the stating point.

In Fig.~\ref{db_vs_nodb}(a), we present the results obtained under the previously mentioned conditions. The blue circles are the numerically-simulated signal acquired using the hybrid NV-DB sensor with our sequence in Fig.~\ref{seq}(b). Conversely, the red triangles depict the signal acquired using an NV without a proximal DB subjected to the sequence shown in the inset of Fig.~\ref{seq}(b). The solid blue and red curves correspond to a standard maximum-likelihood fit to each signal (see subsection ``Maximum likelihood estimation'' in Methods for details). The curves show that the hybrid sensor combined with the sequence in Fig.~\ref{seq}(b)  gives rise to significantly higher contrast compared to the case without DB. 

As it is shown in Fig.~\ref{db_vs_nodb}(b) this superior contrast leads to an accurate estimate of $g$. In particular, with our method we find $g_{\rm est}=1.6(8)$ MHz which is close to the true value $g=1.734$ MHz. Yet, for the single NV sensor (without DB), a maximum-likelihood estimate provides a poorer estimation $g_{\rm est}=2(5)$ MHz. On the other hand, Fig.~\ref{db_vs_nodb}(c)  shows the SNR of both protocols for a growing number of experimental realizations N. To compute the SNR we divide the amplitude of the peak in the power spectrum ($S(f)$) obtained from a signal with no projection noise (see Supplementary Fig.S2) by the standard deviation of the noise. The latter is calculated by subtracting the $S(f)$ computed from the signal without projection noise to the $S(f)$ from the signal with a finite number of measurements. Fig.~\ref{db_vs_nodb}(c) shows that the SNR obtained for the hybrid sensor is $\approx$ 5-6 times higher for  a large range~of~N.

Regarding the impact of distinct noises, we observed that the label $\rm{L}_1$ introduces the most detrimental decoherence channel, leading to the decay of the signal depicted in Fig~{\ref{db_vs_nodb}(a). This decoherence channel affects both protocols (with and without DB) in a similar manner. In addition, the dephasing experienced by the DB (characterized by $T_{\rm 2,DB}=1\ \mu s$) also has a significant impact, which comes into play during the period $2 \tau_2\sim 1\ \mu s$ effectively reducing the signal amplitude obtained from the hybrid sensor. This effect, however, is counteracted by the stronger interactions facilitated by the DB's role in bridging the coupling between NV and ${\rm L_1}$, thus enabling a faster execution of the protocol. 

An extension of the DB dephasing time directly leads to a substantial enhancement in the performance of the hybrid sensor. This improvement could be achieved by incorporating additional $\pi$ pulses during the echo periods in $\tau_2$ (see Fig.~\ref{seq}(b), to identify $\tau_2$ periods), thereby effectively reducing the impact of fast noise. Another potential approach is to explore the utilization of a pair of DBs in a noise-protected state, such as a singlet configuration. Alternatively, surface cleaning techniques could be employed to reduce the impact of the hydrogen bath on the surface, thereby potentially mitigating the noise over the DB. These avenues of research remain open for future investigation, offering promising possibilities for further enhancing the performance of the system. 

    To investigate the impact of varying dephasing times of the DB on the performance of our protocol, we conduct an analysis similar to the one for finding Fig.~\ref{db_vs_nodb}(a) but varying $T_{\rm 2,DB}$ (the rest of the parameters remain unaltered). The resulting maximum-likelihood fits of the obtained signals are presented in Fig.~\ref{T2}(a) for $T_{\rm 2,DB}$ values ranging from $0.5\ \mu$s to $1.5\ \mu$s. Larger dephasing times lead to higher contrasts and more accurate estimates for $g$. For a better illustration, the power spectrum of each signal is shown in Fig.~\ref{T2}(b). In particular, we find that a moderate enhancement from $T_{\rm 2,DB}=1 \ \mu$s to $T_{\rm 2,DB}=1.5 \ \mu$s, leads to an improvement in the estimation of $g$ from $1.6(8)$ MHz to $1.7(5)$ MHz (cf. ``Maximum likelihood estimation'' subsection of Methods). 
\begin{figure}[b]
\includegraphics[width=\linewidth]{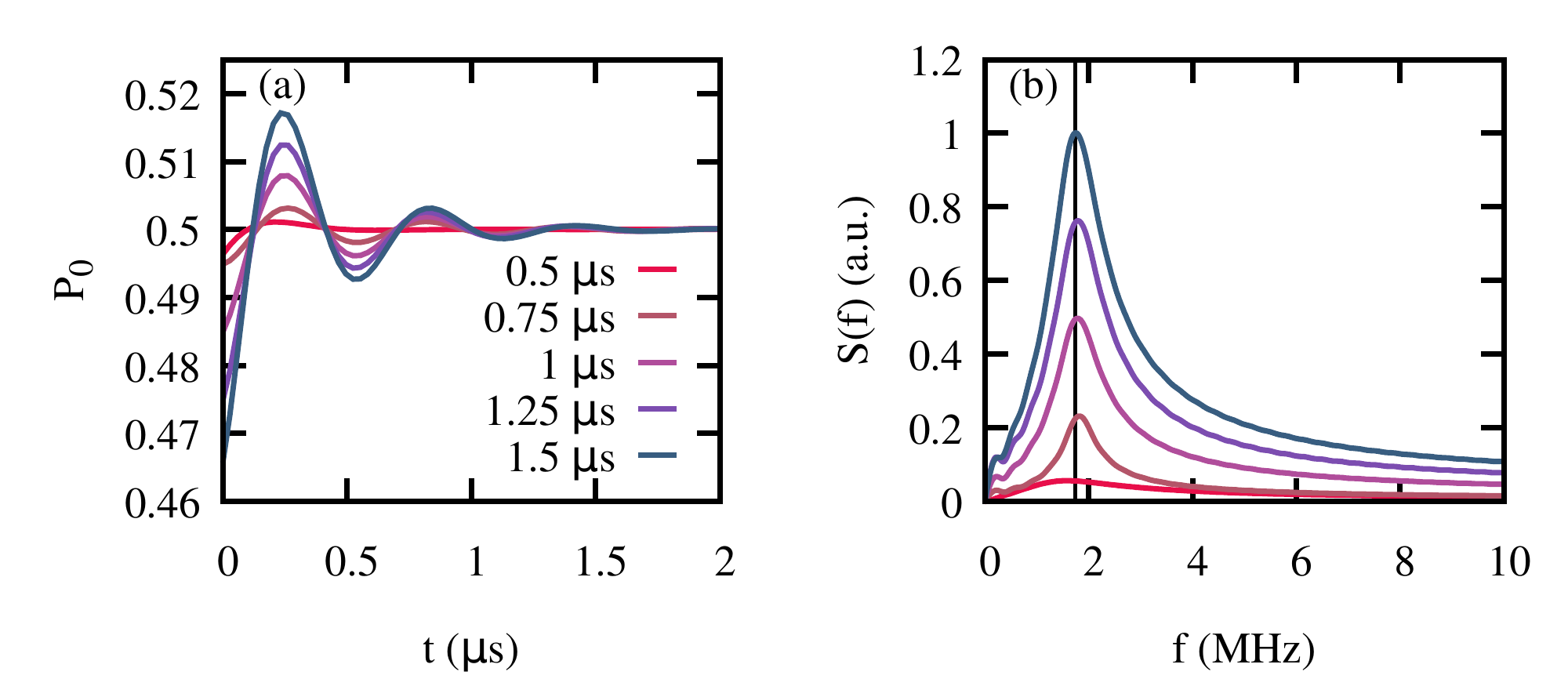}
\caption{{\bf Simulated results for different decoherence times of the DB.} (a) Maximum-likelihood fits of the obtained signals for equivalent parameters as in Fig.~\ref{db_vs_nodb}(a) but varying the dephasing time $T_{\rm 2,DB}$ of the DB as indicated. (b) Power spectrum of the fitted signals, using the same color code as in (a). Reveals the enhanced sensitivity to detect the coupling strength $g=1.734$ MHz, marked with a vertical line, for longer coherence times.}\label{T2}
\end{figure}

To expand upon the obtained results, we investigate a symmetry-preserving configuration, as shown in Fig.~\ref{rad}(a), which allows us to examine the advantages of employing the dangling bond at varying distances. We consider a scenario in which all elements are along the z-axis as illustrated in Fig.~\ref{rad}(a), which we consider to be perpendicular to the diamond surface for convenience. This results in a system with cylindrical symmetry. Then, displacing the DB radially yields results independent of the angle in the xy plane, and we can easily estimate the area in which employing the hybrid sensor is advantageous. The relevant parameters are: $d=8$ nm, $d_1=13$ nm and $d_2=16.5$ nm, yielding $A^{\rm z}_{\rm NV-DB}= (2\uppi)\times0.203\:\rm{MHz}$, $A^{\rm z}_{\rm NV-L_1}=(2\uppi)\times47 \:\rm{kHz}$, $A^{\rm z}_{\rm NV-L_2}=(2\uppi)\times23 \:\rm{kHz}$, $A^{\rm z}_{\rm DB-L_1}=(2\uppi)\times0.831\: \rm{MHz}$, $A^{\rm z}_{\rm DB-L_2}=(2\uppi)\times0.169 \:\rm{MHz}$, $A^{\rm z}_{\rm L_1-L_2}\equiv g=(2\uppi)\times2.423 \:\rm{MHz}$. In Fig.~\ref{rad}(b) it is presented the NV population (in blue) for distinct values of the DB radial deviation after applying our sequence in Fig.~\ref{seq}(b). In red it is shown the case without DB. Recall that the initial configuration pictured in Fig.~\ref{rad}(a) is the most favorable for both cases (with and without DB). Then, the DB is displaced from the z-axis such that the configuration of the hybrid sensor deviates from the ideal, while this is compared with the optimal case for the only NV scenario.

\begin{figure}[t]
\includegraphics[width=\linewidth]{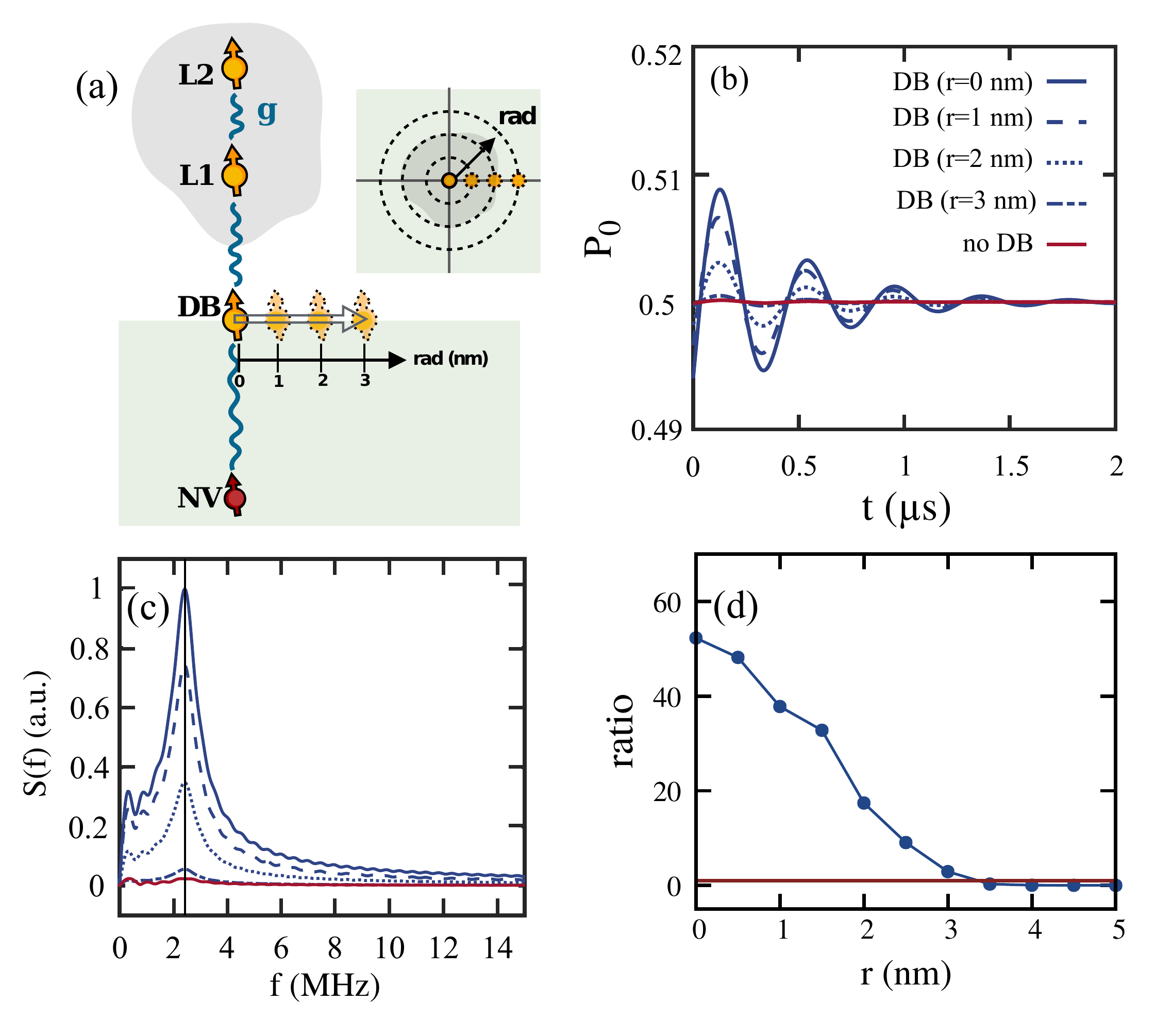}
\caption{{\bf Simulated results for different spatial configurations. }(a) Schematic representation of the hybrid sensor and target configuration. All the elements are aligned along the $z$-axis, while the DB is displaced radially from the center only altering $A^{\rm z}_{\rm NV-DB}$, $A^{\rm z}_{\rm DB-L_1}$ and $A^{\rm z}_{\rm DB-L_2}$, being $A^{\rm z}_{\rm i-j} =\frac{\mu_0 \gamma_{\rm e}^2 \hbar}{4 {\rm \uppi} d_{\rm ij}^3} \left[1 -3 \cos^2{(\theta_{\rm i-j})}\right]$. Inset: Top view of the considered scenario. (b)  In blue the NV ground state population $P_0$ as a function of the DB radial distance resulting from a maximum-likelihood fit, while in red it is shown the case with no DB. (c) Fourier transforms of the signals in (b) with the same color code. The solid vertical line corresponds to $g\equiv\frac{\mu_0 \gamma_{\rm e}^2 \hbar}{4 {\rm \uppi} d_{\rm 12}^3} \left[1 -3 \cos^2{(\theta_{\rm L_1-L_2})}\right]=2.423$ MHz. (d) The ratio between the maxima of the Fourier transform of the hybrid sensor and that of the single NV case, as a function of the radial displacement $r$. The red line is a guide to the eye at~$1$. } \label{rad}
\end{figure}

For the initial configuration the hybrid sensor displays much better sensitivity with a  peak on its Fourier transform which is about 50 times higher than the case without DB, see Fig.~\ref{rad}(c). When the DB is moved away from the ideal configuration, the hybrid sensor still exhibits superiority up to a radius above $3$ nm. Fig.~\ref{rad}(d) illustrates how the height ratio of the Fourier transform changes with respect to the radius for the sequence with and without DB. It is worth noting that the sequence with DB consistently exhibits a stronger signal quality compared to the sequence without DB, until the radius reaches approximately 3 nm.
If the DB is located further away, the amplitude of the couplings involving the DB keep being greater (the term amplitude refers to the angle-independent component of the coupling which, for instance, at r=$3.5$ nm have a value of $A^{\rm 0}_{\rm NV-DB} =\frac{\mu_0 \gamma_{\rm e}^2 \hbar}{4 \uppi d^3}=(2\uppi)\times78$ kHz and $A^{\rm 0}_{\rm DB-L_1} = (2\uppi)\times155$ kHz, whereas the direct NV-L${_1}$ coupling is $A^{\rm 0}_{\rm NV-L_1}=(2\uppi)\times24$ kHz. Thus, in terms of distances the DB could be placed further away and still achieve a stronger signal with the hybrid sensor. However, as the DB is moved outwards, the angles $\theta_{\rm NV-DB}$ and $\theta_{\rm DB-L_1}$ approach the magic angle, nullifying the dipolar interactions that involve the DB. Then, in this specific symmetric scenario these angles limit the area in which using the DB is favorable up to $\approx 30 \ \text{nm}^2$. Note that according to experimental studies~\cite{myers2014probing} it is likely to have a DB in this area. Hence, this leads to multiple configurations in which the hybrid sensor driven by our sequence in Fig.~\ref{seq}.(b) outperforms the NV-only sensor. 

\section{Conclusions}
In summary, our work introduces a  detection protocol utilising a hybrid sensor and adequately designed MW sequence that exploits the presence of dangling bonds. With this, we demonstrate the encoding of an inter-label coupling constant in oscillations of the NV fluorescence oscillations, while decoherence effects are mitigated. We showed the superior signal-to-noise ratio achieved by our method compared to the standard scenario without a DB. This research paves the way for new possibilities in quantum sensing with solid-sate defects through the utilization of nanoscale hybrid sensors as it is the case of several DBs taken to decoherence free subspaces.

\section{Methods}
\subsection{Implementation of decoherence}
We employ a Master equation to introduce the distinct decoherence channels in our numerical models. In particular, we adopt the following expression:
\begin{align}
    \frac{d\rho}{dt}=-{\rm i}\left[H,\rho\right]+\sum_{i}^N L_i, \label{eq:Linbland}
\end{align}
where $\rho$ is the density matrix of the system, $H$ is the Hamiltonian and $L_i$ are operators that describe the effect of irreversible processes on the $i$-th spin. More specifically,
\begin{align}
L_{\rm i}=&\frac{1}{4T_{\rm 2,i}}\left(2\sigma_{\rm i}^{\rm z}\rho \sigma_{\rm i}^{\rm z} - 2\rho\right)\nonumber\\
+&\Gamma_{\rm i}\left[\left(m_{\rm i}+1\right)\left(2\sigma_{\rm i}^+\rho\sigma_{\rm i}^- - \sigma_{\rm i}^+\sigma_{\rm i}^+\rho - \rho\sigma_{\rm i}^-\sigma_{\rm i}^+\right)\right.\nonumber\\
+&\left.m_{\rm i}\left(2\sigma_{\rm i}^-\rho\sigma_{\rm i}^+ - \sigma_{\rm i}^-\sigma_{\rm i}^+\rho - \rho\sigma_{\rm i}^+\sigma_{\rm i}^-\right)\right] \label{eq:L}
\end{align}
where  $\sigma_{\pm}$ are ladder operators defined as $\sigma_{+}=\sigma_{\rm x}+{\rm i}\sigma_{\rm y}$ and $\sigma_{-}=\sigma_{\rm x}-{\rm i}\sigma_{\rm y}$, $\Gamma=\frac{1}{2(2m+1)T_1}$ and $m=\exp\left(-\frac{2\uppi \hbar g_{\rm e} B_{\rm z}}{K_{\rm B} T}\right)$. The first line introduces the effects of dephasing, while the second and third lines effectively model the relaxation.

\subsection{Cross-talk}
Deviations to single addressing lead to potential undesired effects. In particular, a system that  comprises a number $N$ of 1/2 spins under the action of two simultaneous drivings is described by

\begin{align}
    H=&\sum_{{\rm k}}^{N}  \frac{\omega_{\rm k} \sigma_{\rm k}^{\rm (z)}}{2}+\Omega_1\sum_{\rm k}^{N}\sigma_{\rm k}^{\rm (x)} \cos\left(\tilde{\omega}_1\:t+\phi_1\right) \nonumber\\
    +&\Omega_2\sum_{\rm k}^{N}\sigma_{\rm k}^{\rm (x)}\cos \left(\tilde{\omega}_2\:t+\phi_2\right),
\end{align}
where $\omega_{\rm k}$ corresponds to the Larmor frequency of the k-th element, $\sigma_{\rm k}^{\rm (x,y,z)}$ to the Pauli matrices,
and $\phi_1$ and $\phi_2$ are the pulse phases which can take a value of $0$ or $\uppi/2$ (corresponding to X and Y pulses, respectively).

In a rotating frame with respect to $H_0=\sum_{\rm k}^{N}  \frac{\omega_k \sigma_{\rm k}^{\rm (z)}}{2}$, one finds 

\begin{align}
    H_{\rm{CT}}=&\Omega_1\sum_{\rm k}^{N}\left[\rm{e}^{i\left(\delta_{\rm k}^1 t + \phi_1\right)}\sigma_{\rm k}^{(-)}+\rm{H.c.}\right]\nonumber\\
    +&\Omega_2\sum_{\rm k}^{N}\left[{\rm e}^{{\rm i}\left(\delta_{\rm k}^2 t + \phi_2\right)}\sigma_{\rm k}^{(-)}+\rm{H.c.}\right],\label{eq:ct}
\end{align}
with $\delta_{\rm k}^1=\omega_{\rm k}-\tilde{\omega}_1$ and $\delta_{\rm k}^2=\omega_{\rm k} - \tilde{\omega}_2$ being the detunings of the k-th element with respect to the first and second driving.

Notice that, if the detunings $\delta_{\rm k}^1$ and $\delta_{\rm k}^2$ are much greater than the amplitudes $\Omega_1$ and $\Omega_2$, the RWA can be invoked, and only the terms that are intentionally addressed (i.e., those with $\delta_{\rm k}^{\rm j} = 0$ )  survive. However, in case $\delta_{\rm k}^{\rm j}$ is comparable with the Rabi frequencies, one cannot remove time dependent terms as they introduce pulse deviations. 

The numerical simulations shown in the main text are performed from Eq.~\ref{eq:Linbland} where H is the interaction Hamiltonian \ref{eq:ZZ} during the free evolution periods and, during the pulses, the latter plus $H_{CT}$ . Recall that N is the total number of elements playing a role in the sequence, this is, 4 in the first sequence (NV, DB, ${\rm L_1}$ and ${\rm L_2}$) and 3 in the second (NV, ${\rm L_1}$ and ${\rm L_2}$).

\subsection{Maximum likelihood estimation}
In the following, we explain the maximum-likelihood estimate employed to fit the noisy signals, as well as to provide an estimate for the coupling strength $g$. Let us consider $\{y_{\rm n}\}$ to denote a set of observations with uncertainty $\{\Delta y_{\rm n}\}$ at their corresponding times $\{t_{\rm n}\}$ for $n=0,\ldots, M$. We propose a model $y_{\rm f}(t;{\bf p})$ to fit the observations $\{y_{\rm n}\}$ where ${\bf p}$ represents the free parameters of the model. Then, we can compute the likelihood $L({\bf p})$ that such model $y_{\rm f}(t;{\bf p})$ explains the observations $\{y_{\rm n}\}$. For that, we assume normally distributed observations, which is justified in our case given the large amount of measurements per time instant. Hence, the likelihood reads
\begin{align}
L({\bf p})=\Pi_{\rm n=0}^M \frac{1}{\Delta y_{\rm n}\sqrt{2\uppi}}{\rm e}^{-(y_{\rm n}-y_{\rm f}(t_n;{\bf p}))^2/(2\Delta y_{\rm n}^2)}.
\end{align}
The maximum-likelihood estimate ${\bf p}_{\rm MLE}$ follows from the parameters that maximize the likelihood, i.e. ${\bf p}_{\rm MLE}={\rm max}_{{\bf p}}L({\bf p})$. This is similar to the procedure in Bayesian estimation theory given uniform priors for ${\bf p}$. The uncertainty associated with this estimate can be easily obtained from $\frac{1}{\sigma^2_{\rm j}}=-d^2 \tilde{L}({\bf p})/d p^2_{\rm j}|_{{\bf p}={\bf p}_{\rm MLE}}$ where $\tilde{L}({\bf p}_{\rm MLE})=1$ is a normalized likelihood. Note that since $L({\bf p})$ is a differentiable function, $dL({\bf p})/d p_j|_{{\bf p}_{\rm MLE}}=0$ by construction. 
In our case, we consider a phenomenological model  
\begin{align}\label{eq:ymodel}
y(t;{\bf p})=\frac{1}{2}+p_1 e^{p_2 t}\cos(2\pi p_3 t+p_4),
\end{align}
to account for the noisy signal. See Supplementary Note 2 for details on the agreement between the exact numerical results and this phenomenological model. In this manner,  there are four free parameters, ${\bf p}=(p_1,p_2,p_3,p_4)$. The expression given in Eq.~\eqref{eq:ymodel} is the expected behavior of the signal under decoherence (with a characteristic time $-1/p_2$) and from where the identification $p_3=g$ is direct (see main text). In addition, the parameters $p_1$ and $p_4$ are included to account for other imperfections. The solid curves presented in the main text stem from $y(t;{\bf p}_{\rm MLE})$ to each of the cases considered in the main text. 

\section{data availability}
The complete dataset supporting the findings of this work is accessible. On the one hand, secondary data reused and/or analyzed from other studies is specified along the text, in case any doubts or inquires, please contact the corresponding authors at jcasanovamar@gmail.com. On the other hand, original data generated in this study (via numerical simulations) are available upon request, kindly direct any further correspondence to the aforementioned email address.

\section{code availability}
The source code and associated materials used in this research are available upon request. For access to the code, inquiries, or additional information, please contact the corresponding authors at jcasanovamar@gmail.com. We are committed to facilitating transparency and collaboration in our research and will provide assistance and code access as needed.

\begin{acknowledgements}
A.~B.~U. and P.~A.~B. acknowledge the financial support of the IKUR STRATEGY (IKUR-IKA-23/22) and (IKUR-IKA-23/04), respectively. C. M.-J. acknowledges the predoctoral MICINN Grant No. PRE2019-088519. J.~C. acknowledges the Ram\'{o}n y Cajal  (RYC2018-025197-I) research fellowship, the financial support from Spanish Government via the Nanoscale NMR and complex systems (PID2021-126694NB-C21) project, the ELKARTEK project Dispositivos en Tecnolog\'i{a}s Cu\'{a}nticas (KK-2022/00062), and the Basque Government grant IT1470-22. 
\end{acknowledgements}

\section{Author Contributions}

A.~B.~U. and P.~A.~B. equally contributed to the work by developing the theoretical expressions, performing the numerical simulations and writing the paper. C. M. J. contributed to the development of the idea and discussions. R. P. performed the statistical analysis of the numerical simulations. J. C. supervised the process and provided the theoretical background and research direction. All authors reviewed the manuscript.

\section{Competing Interests}
The authors declare no competing interests.

\clearpage
\widetext
\begin{center}
\textbf{ \large Supplemental Material: \\ High-Field Microscale NMR with Nitrogen-Vacancy Centers for Dipolar Coupled Systems}
\end{center}
\setcounter{equation}{0} \setcounter{figure}{0} \setcounter{table}{0}
\setcounter{page}{1} \makeatletter \global\long\def\theequation{S\arabic{equation}}
 \global\long\def\thefigure{S\arabic{figure}}
 \global\long\def\bibnumfmt#1{[S#1]}
 \global\long\def\citenumfont#1{S#1}

\title{Supplemental Material of Amplified Nanoscale Detection of Labelled Molecules via Surface Electrons on Diamond}
\author{A. Biteri-Uribarren}
\thanks{These authors have equally contributed to this work.}
\affiliation{Department of Physical Chemistry, University of the Basque Country UPV/EHU, Apartado 644, 48080 Bilbao, Spain}
\affiliation{EHU Quantum Center, University of the Basque Country UPV/EHU, Leioa, Spain}

\author{P. Alsina-Bol\'{i}var}
\thanks{These authors have equally contributed to this work.}
\affiliation{Department of Physical Chemistry, University of the Basque Country UPV/EHU, Apartado 644, 48080 Bilbao, Spain}
\affiliation{EHU Quantum Center, University of the Basque Country UPV/EHU, Leioa, Spain}

\author{C. Munuera-Javaloy}
\affiliation{Department of Physical Chemistry, University of the Basque Country UPV/EHU, Apartado 644, 48080 Bilbao, Spain}
\affiliation{EHU Quantum Center, University of the Basque Country UPV/EHU, Leioa, Spain}
\author{R. Puebla}
\affiliation{Department of Physics, University Carlos III of Madrid, Avda. de la Universidad 30, 28911  Legan{\'e}s, Madrid, Spain}
\author{J. Casanova}
\affiliation{Department of Physical Chemistry, University of the Basque Country UPV/EHU, Apartado 644, 48080 Bilbao, Spain}
\affiliation{EHU Quantum Center, University of the Basque Country UPV/EHU, Leioa, Spain}
\affiliation{IKERBASQUE,  Basque  Foundation  for  Science, Plaza Euskadi 5, 48009 Bilbao,  Spain}
\affiliation{Corresponding author: jcasanovamar@gmail.com}

\maketitle

\setcounter{equation}{0} \setcounter{figure}{0} \setcounter{table}{0}
\setcounter{page}{1} \makeatletter \global\long\def\theequation{S\arabic{equation}}
 \global\long\def\thefigure{S\arabic{figure}}
 \global\long\def\bibnumfmt#1{[S#1]}
 \global\long\def\citenumfont#1{S#1}

\section{Supplementary Note 1: CONFIGURATIONS AND COUPLINGS}\label{app:conf}

The performance of the sequences depicted in Fig.1 (b) and (c) of the main text can vary significantly depending on the angle between the joining vector of the two elements and the external magnetic field (${\bf B}=B_0\hat{\bf z}$) due to the strong dependence of the dipolar couplings with this angle. To ensure a fair comparison between the two sequences, it is important to choose the positions of the elements such that neither sequence is favored. Thus, in our case, the angles $\theta_{\rm NV-DB}$ and $\theta_{\rm DB-L_1}$ are similar to $\theta_{\rm NV-L_1}$ so that the difference in the coupling amplitudes ($A^{\rm z}_{\rm NV-DB}$, $A^{\rm z}_{\rm DB-L_1}$, and $A^{\rm z}_{\rm NV-L_1}$) depends primarily on the distances $d$, $d_{\rm DB-L_1}$, and $d_1$. Supplemental Fig. \ref{conf} is a generic scheme of the system, where the relevant angles and vectors are illustrated.

\begin{figure}[h]
\includegraphics[width=0.5\linewidth]{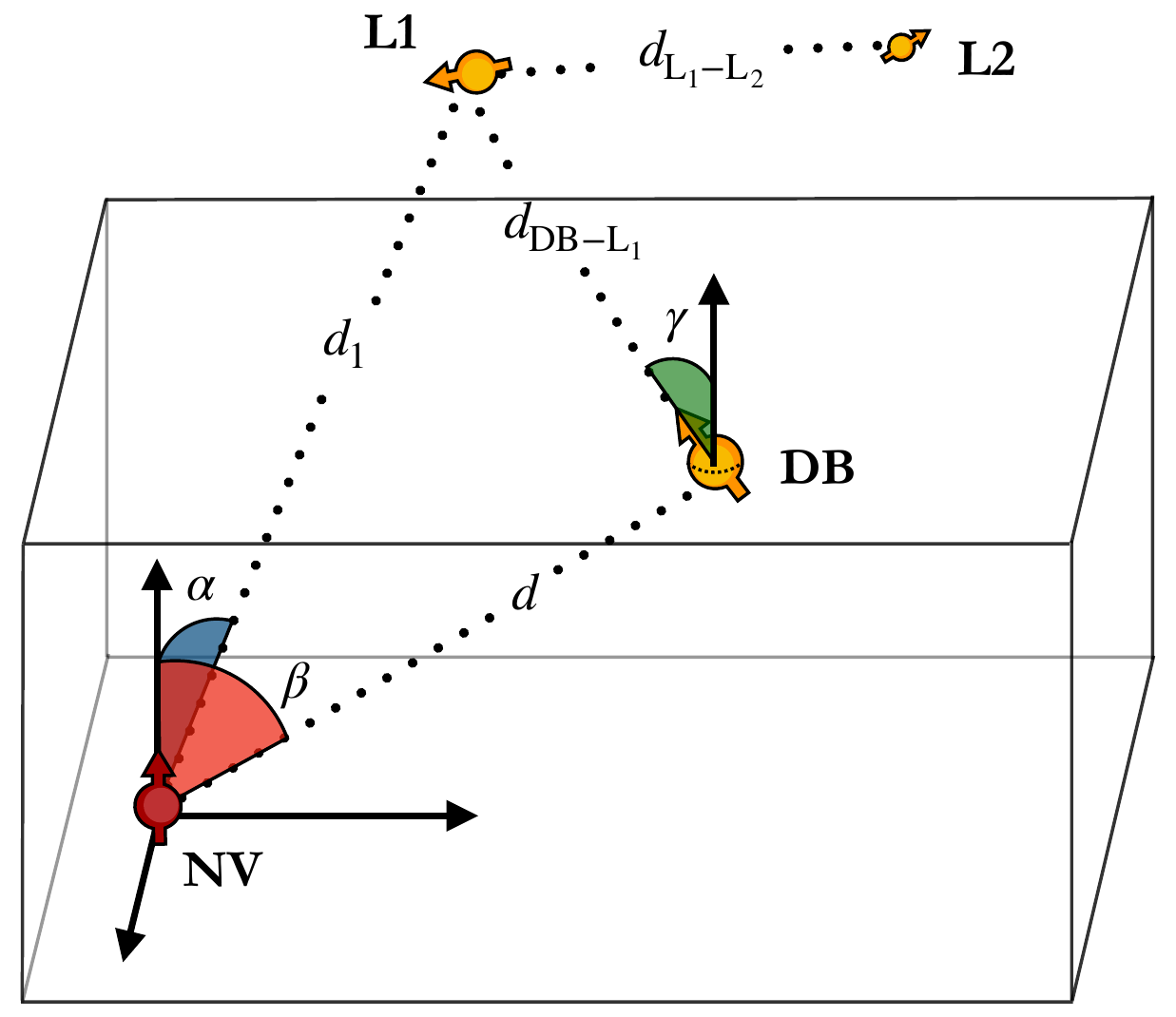}
\caption{ Scheme of the configuration (not to scale). For the numerical simulations in the main text, the relevant parameters have the values: $\alpha\equiv\theta_{\rm NV-L_1}=13.4^{\circ}$, $\beta\equiv\theta_{\rm NV-DB}=13.8^{\circ}$ and $\gamma\equiv\theta_{\rm DB-L_1}=13.5^{\circ}$, with distances $d=5.6 \rm{nm}$, $d_{\rm DB-L_1}=5.7\rm{nm}$ and $d_1=11.3\rm{nm}$.} \label{conf}
\end{figure}
\newpage

\section{Supplementary note 2: PROJECTION NOISE FIT AND EXPECTED VALUE}\label{app:fitmodel}
In this section we present the data with no projection noise, i.e. the clean signal on top of which we simulate the experiments that are in Fig.2 of the main text. The core of our claim resides in this data-set: the signal extracted employing the protocol with DB results in a six-fold contrast improvement with respect to the single NV case, what leads to, when we dress it with experimental noise, better resolution and higher SNR compared to the alternative no-DB scenario.

Besides, Supplemental Fig.~\ref{expected} shows that the heuristic model chosen to perform the maximum-likelihood estimation, cf. Eq. $11$ in the main text, indeed reproduces the original signal, thus supporting its use to extract the relevant coupling strength $g$.

\begin{figure*}[h]
\includegraphics[width=0.75\linewidth]{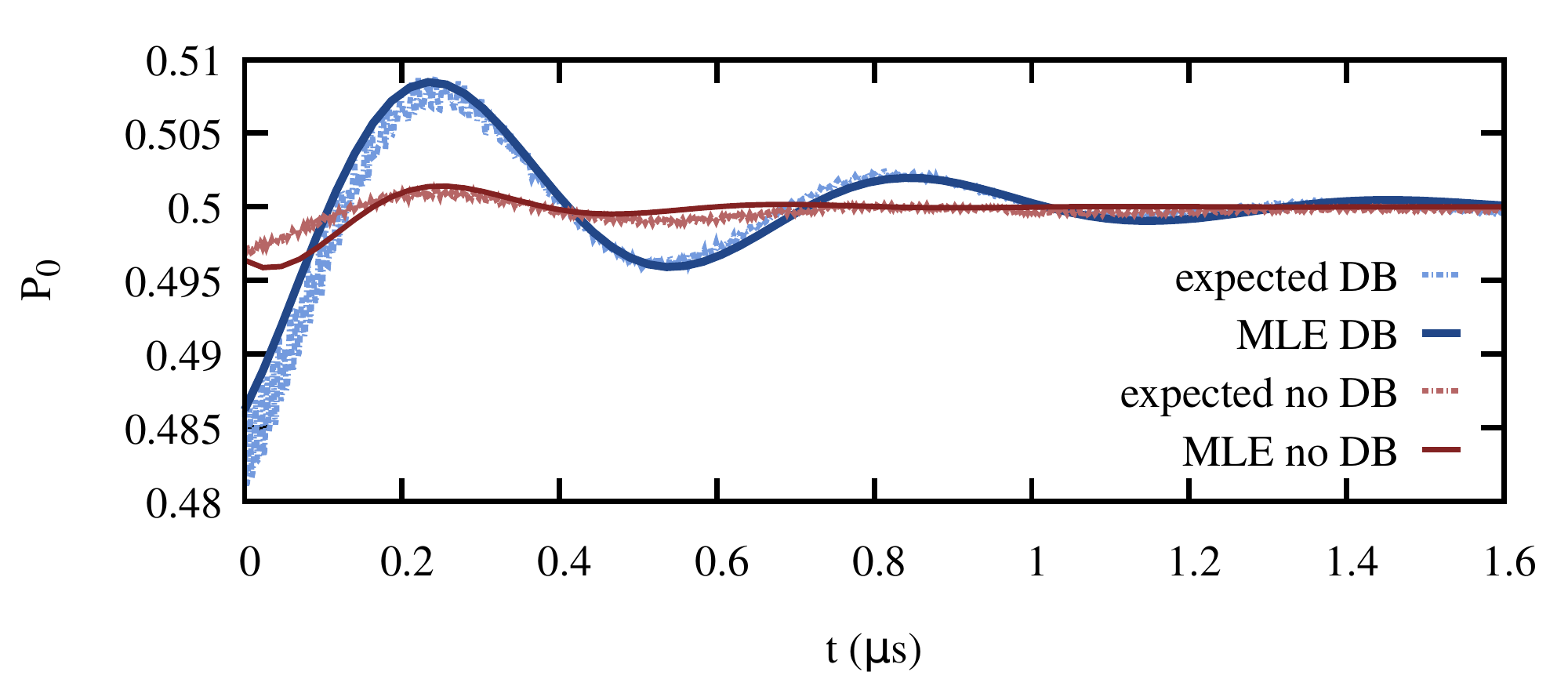}
\caption{The dashed blue (red) line represents the expected NV population evolution without projection noise for the hybrid NV-DB (only-NV) sensor. The continuous lines, in the corresponding colors, are the maximum-likelihood estimate with the phenomenological model of Eq. $11$ in the ``Maximum likelihood estimation'' subsection of Methods .} \label{expected}
\end{figure*}

\end{document}